\documentclass[letterpaper,twocolumn,superscriptaddress,floatfix]{revtex4}

\usepackage[latin1]{inputenc}
\usepackage{bm}
\usepackage{multirow,amssymb,amsbsy,amsmath}
\usepackage{graphicx}
\usepackage{verbatim}
\usepackage{braket}
\makeatletter
\usepackage{pifont}
\makeatother

\usepackage{hyperref}
\hypersetup{hidelinks,colorlinks=true,allcolors=black,pdfstartview=Fit,breaklinks=true}
\usepackage{dcolumn}
\usepackage{ifthen}
\usepackage{booktabs}
\usepackage{nicefrac}
\usepackage{float}
\usepackage{color,xcolor}
\usepackage{upgreek}

\begin{document}

\title{Nanoscale positioning and in-situ enhancement of single G center in silicon using a fluorescence-localization technique}
\affiliation{CAS Key Laboratory of Quantum Information, University of Science and Technology of China, Hefei, Anhui 230026, China}
\affiliation{Anhui Province Key Laboratory of Quantum Network, University of Science and Technology of China, Hefei, Anhui 230026, China}
\affiliation{CAS Center For Excellence in Quantum Information and Quantum Physics, University of Science and Technology of China, Hefei, 230026, China}
\affiliation{Hefei National Laboratory, University of Science and Technology of China, Hefei 230088, China}

\author{Yu-Hang Ma}
\thanks{These authors contributed equally to this work.}
\affiliation{CAS Key Laboratory of Quantum Information, University of Science and Technology of China, Hefei, Anhui 230026, China}
\affiliation{Anhui Province Key Laboratory of Quantum Network, University of Science and Technology of China, Hefei, Anhui 230026, China}
\affiliation{CAS Center For Excellence in Quantum Information and Quantum Physics, University of Science and Technology of China, Hefei, 230026, China}

\author{Nai-Jie Guo}
\thanks{These authors contributed equally to this work.}
\affiliation{CAS Key Laboratory of Quantum Information, University of Science and Technology of China, Hefei, Anhui 230026, China}
\affiliation{Anhui Province Key Laboratory of Quantum Network, University of Science and Technology of China, Hefei, Anhui 230026, China}
\affiliation{CAS Center For Excellence in Quantum Information and Quantum Physics, University of Science and Technology of China, Hefei, 230026, China}
\affiliation{Hefei National Laboratory, University of Science and Technology of China, Hefei 230088, China}

\author{Wei Liu}
\thanks{These authors contributed equally to this work.}
\affiliation{CAS Key Laboratory of Quantum Information, University of Science and Technology of China, Hefei, Anhui 230026, China}
\affiliation{Anhui Province Key Laboratory of Quantum Network, University of Science and Technology of China, Hefei, Anhui 230026, China}
\affiliation{CAS Center For Excellence in Quantum Information and Quantum Physics, University of Science and Technology of China, Hefei, 230026, China}

\author{Xiao-Dong Zeng}
\affiliation{CAS Key Laboratory of Quantum Information, University of Science and Technology of China, Hefei, Anhui 230026, China}
\affiliation{Anhui Province Key Laboratory of Quantum Network, University of Science and Technology of China, Hefei, Anhui 230026, China}
\affiliation{CAS Center For Excellence in Quantum Information and Quantum Physics, University of Science and Technology of China, Hefei, 230026, China}

\author{Lin-Ke Xie}
\affiliation{CAS Key Laboratory of Quantum Information, University of Science and Technology of China, Hefei, Anhui 230026, China}
\affiliation{Anhui Province Key Laboratory of Quantum Network, University of Science and Technology of China, Hefei, Anhui 230026, China}
\affiliation{CAS Center For Excellence in Quantum Information and Quantum Physics, University of Science and Technology of China, Hefei, 230026, China}

\author{Jun-You Liu}
\affiliation{CAS Key Laboratory of Quantum Information, University of Science and Technology of China, Hefei, Anhui 230026, China}
\affiliation{Anhui Province Key Laboratory of Quantum Network, University of Science and Technology of China, Hefei, Anhui 230026, China}
\affiliation{CAS Center For Excellence in Quantum Information and Quantum Physics, University of Science and Technology of China, Hefei, 230026, China}
\affiliation{Hefei National Laboratory, University of Science and Technology of China, Hefei 230088, China}

\author{Ya-Qi Wu}
\affiliation{CAS Key Laboratory of Quantum Information, University of Science and Technology of China, Hefei, Anhui 230026, China}
\affiliation{Anhui Province Key Laboratory of Quantum Network, University of Science and Technology of China, Hefei, Anhui 230026, China}
\affiliation{CAS Center For Excellence in Quantum Information and Quantum Physics, University of Science and Technology of China, Hefei, 230026, China}

\author{Yi-Tao Wang}
\affiliation{CAS Key Laboratory of Quantum Information, University of Science and Technology of China, Hefei, Anhui 230026, China}
\affiliation{Anhui Province Key Laboratory of Quantum Network, University of Science and Technology of China, Hefei, Anhui 230026, China}
\affiliation{CAS Center For Excellence in Quantum Information and Quantum Physics, University of Science and Technology of China, Hefei, 230026, China}

\author{Zhao-An Wang}
\affiliation{CAS Key Laboratory of Quantum Information, University of Science and Technology of China, Hefei, Anhui 230026, China}
\affiliation{Anhui Province Key Laboratory of Quantum Network, University of Science and Technology of China, Hefei, Anhui 230026, China}
\affiliation{CAS Center For Excellence in Quantum Information and Quantum Physics, University of Science and Technology of China, Hefei, 230026, China}

\author{Jia-Ming Ren}
\affiliation{CAS Key Laboratory of Quantum Information, University of Science and Technology of China, Hefei, Anhui 230026, China}
\affiliation{Anhui Province Key Laboratory of Quantum Network, University of Science and Technology of China, Hefei, Anhui 230026, China}
\affiliation{CAS Center For Excellence in Quantum Information and Quantum Physics, University of Science and Technology of China, Hefei, 230026, China}

\author{Chun Ao}
\affiliation{CAS Key Laboratory of Quantum Information, University of Science and Technology of China, Hefei, Anhui 230026, China}
\affiliation{Anhui Province Key Laboratory of Quantum Network, University of Science and Technology of China, Hefei, Anhui 230026, China}
\affiliation{CAS Center For Excellence in Quantum Information and Quantum Physics, University of Science and Technology of China, Hefei, 230026, China}

\author{Haifei Lu}
\affiliation{School of Physics and Mechanics, Wuhan University of Technology, Wuhan 430070, China}

\author{Jian-Shun Tang}
\email{tjs@ustc.edu.cn}
\affiliation{CAS Key Laboratory of Quantum Information, University of Science and Technology of China, Hefei, Anhui 230026, China}
\affiliation{Anhui Province Key Laboratory of Quantum Network, University of Science and Technology of China, Hefei, Anhui 230026, China}
\affiliation{CAS Center For Excellence in Quantum Information and Quantum Physics, University of Science and Technology of China, Hefei, 230026, China}
\affiliation{Hefei National Laboratory, University of Science and Technology of China, Hefei 230088, China}

\author{Chuan-Feng Li}
\email{cfli@ustc.edu.cn}
\affiliation{CAS Key Laboratory of Quantum Information, University of Science and Technology of China, Hefei, Anhui 230026, China}
\affiliation{Anhui Province Key Laboratory of Quantum Network, University of Science and Technology of China, Hefei, Anhui 230026, China}
\affiliation{CAS Center For Excellence in Quantum Information and Quantum Physics, University of Science and Technology of China, Hefei, 230026, China}
\affiliation{Hefei National Laboratory, University of Science and Technology of China, Hefei 230088, China}

\author{Guang-Can Guo}
\affiliation{CAS Key Laboratory of Quantum Information, University of Science and Technology of China, Hefei, Anhui 230026, China}
\affiliation{Anhui Province Key Laboratory of Quantum Network, University of Science and Technology of China, Hefei, Anhui 230026, China}
\affiliation{CAS Center For Excellence in Quantum Information and Quantum Physics, University of Science and Technology of China, Hefei, 230026, China}
\affiliation{Hefei National Laboratory, University of Science and Technology of China, Hefei 230088, China}

\date{\today }
\begin{abstract}

Silicon-based semiconductor nanofabrication technology has achieved a remarkable level of sophistication and maturity, and color centers in silicon naturally inherit this advantage. Besides, their emissions appear in telecommunication bands, which makes them play a crucial role in the construction of quantum network. To address the challenge of weak spontaneous emission, different optical cavities are fabricated to enhance the emission rate. However, the relative location between cavity and emitter is random, which greatly reduce the success probability of enhancement. Here, we report on a fluorescence-localization technique (FLT) for precisely locating single G center in silicon and embedding it in the center of a circular Bragg grating cavity in situ, achieving 240-times improvement of the success probability. We observe a 30-fold enhancement in luminescence intensity, 2.5-fold acceleration of the emission from single G center, corresponding to a Purcell factor exceeding 11. Our findings pave the way for the large-scale integration of quantum light sources including those with spins.

\end{abstract}

\maketitle
\bibliographystyle{prsty}

\section*{Introduction}
Quantum information technology is poised to have a profound impact on human society \cite{Bennett2000}. The development of quantum networks presents new opportunities and challenges including quantum communication \cite{Gisin2007}, quantum computing \cite{Nielsen2010}, and quantum sensing \cite{Degen2017}. Single-photon emitter (SPE) is one of the central building blocks for many leading quantum information technologies, and a multitude of promising material systems have been developed. In particular, solid-state SPEs based on atom-like emitters, such as fluorescent atomic defects, attract huge attention due to the potential to combine the exceptional optical properties of atoms with the convenience and scalability of a solid-state host system. Widely studied SPEs include the nitrogen-vacancy (NV) center and silicon-vacancy (SiV) center in diamond \cite{Doherty2013,Hepp2014}, the divacancy defect and the silicon vacancy defect in silicon carbide \cite{Lohrmann2017,Wang2019}, and carbon-related defect in hexagonal boron nitride (hBN) \cite{Guo2023}.

Silicon is a promising material system owing to its exceptional scalability and potential for integration with established large-scale silicon photonics platforms \cite{Silverstone2016,Samkharadze2018}. It is directly compatible with modern mature semiconductor processes. There are primarily two technical routes for generating single photon sources in silicon. The first route involves creating emitters using rare earth element dopants \cite{Yin2013,Raha2020,Chen2020,Ourari2023}. The second route entails creating isolated defects directly in the silicon that function as artificial atoms \cite{Durand2021,Khoury2022}, which is adopted in this work. Recently, a variety of color centers in silicon have been isolated, including G center \cite{Hollenbach2020,Redjem2020,Udvarhelyi2021,Baron2022_1,Hollenbach2022_1,Prabhu2023}, T center \cite{Higginbottom2022} and W center \cite{Hollenbach2022_1,Baron2022_2}. These color centers exhibit a significant shared advantage: their zero-phonon line (ZPL) emission naturally resides within the telecommunication wavelength range, which enables efficient long-distance quantum communication without frequency conversion processes. The drawback of these color centers which constrains their practical applications is their inherently weak spontaneous emission. To address this limitation, people developed strategies to integrate these centers into nanophotonic structures, which significantly enhance their emission efficiency. Recent advancements have demonstrated successful integration with various photonic platforms, including photonic crystal cavities \cite{Johnston2024,Islam2024,Redjem2023,Saggio2024,Komza2024}, microring resonators \cite{Tait2020,Lefaucher2023}, metasurfaces \cite{Zhu2020}, nanopillars \cite{Hollenbach2022_2} and circular Bragg grating (CBG) cavity \cite{Lefaucher2024}, all of which have shown substantial enhancement in ZPL emission intensity. However, in these works, the relative locations between the cavities and single defect centers are random, which makes the success probabilities of emission enhancement quite low; or else, the large ensembles of defects are just used instead of single emitters.

\begin{figure*}[htbp]
  \centering
  \includegraphics[width=0.8\textwidth]{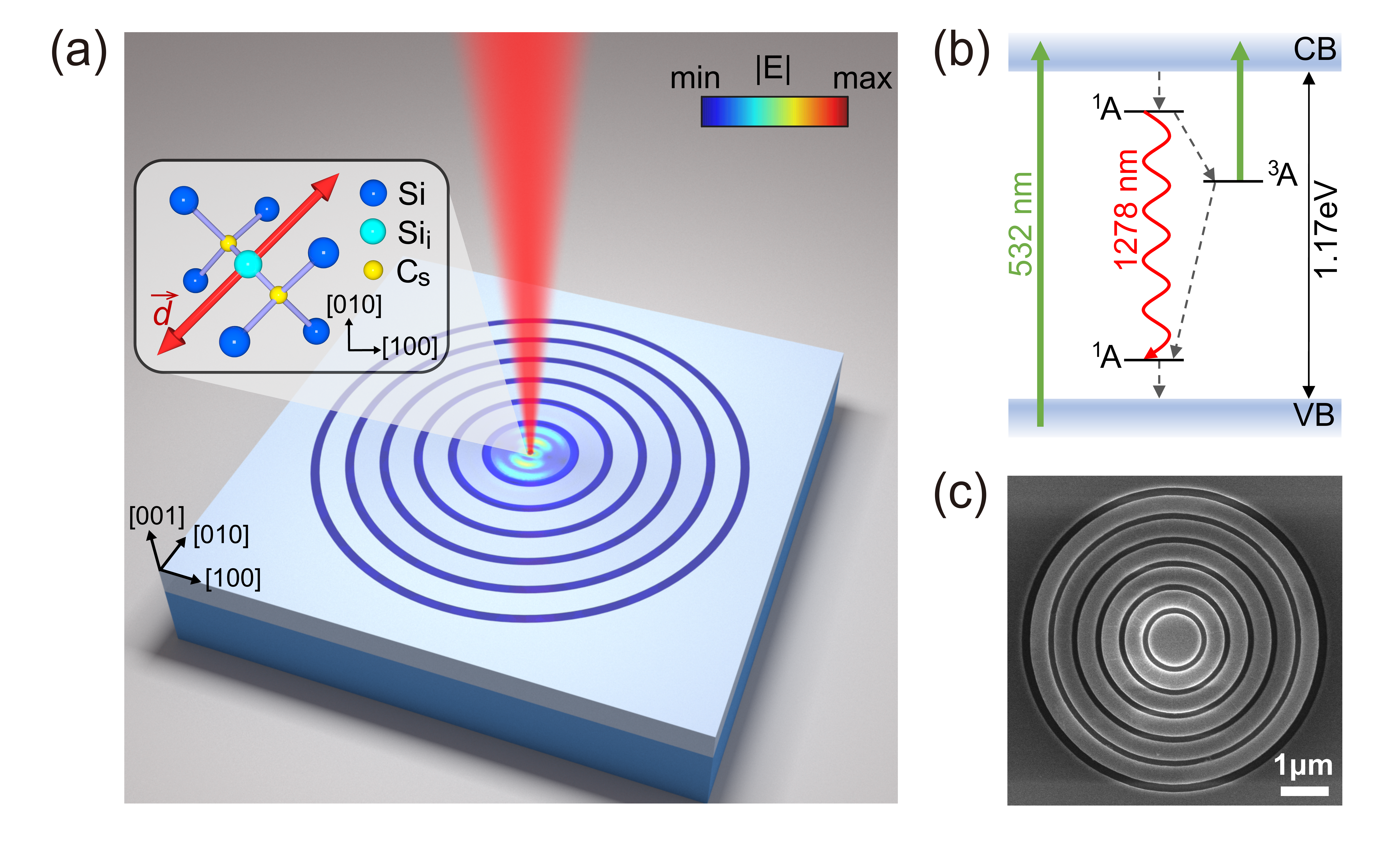}
  \caption{\textbf{Schematic of a single G center embedded in a CBG cavity.} (a) A schematic of a CBG cavity on SOI generating collimated emission into free space. The inset shows a schematic of the G center in the silicon lattice, which consists of a pair of substituted carbon atoms (yellow) and an interstitial silicon atom (cyan). The red arrow represents the direction of a single G center dipole. (b) Energy level diagram of G center in silicon including a ground singlet state, an excited singlet state and a metastable triplet state. (c) Scanning electron microscope (SEM) image of the CBG cavity.}\label{fig1}
\end{figure*}

In our work, to address this problem, we develop and implement an advanced fluorescence-localization technique (FLT) that enables accurate positioning and alignment of the quantum emitters within the photonic structure with $\sim$15 nm spatial accuracy, creating a high success probability of enhancement of 60\%, 240 times of that of random match. Through precise spectral alignment of the cavity resonance with the ZPL of individual G centers, we achieved a remarkable 30-fold enhancement in luminescence intensity accompanied by a 2.5-fold acceleration of the single-photon emission rate, corresponding to a Purcell factor exceeding 11.

Here we take CBG cavity as the example, which is formed by etching concentric rings of equal spacing on silicon. It is considered as a great nanophotonic structure for its advantages of broadband Purcell enhancement and efficient light extraction and collection through far-field radiation modes. Furthermore, the inherent symmetry of the CBG cavity design provides considerable tolerance to dipole orientation, allowing for enhancement of individually oriented dipoles \cite{Rickert2019}. This cavity has previously been widely used in other platforms, including color centers in diamond \cite{Li2015}, defects in hBN \cite{Froch2021}, quantum dots \cite{Liu2019,Kolatschek2021}, and W center ensemble in silicon \cite{Lefaucher2024}.

\section*{Results}
\textbf{Device design. }The device that we report features a single G center deterministically embedded in a silicon-based CBG cavity, as illustrated in Fig. 1(a). According to the density functional theory \cite{Udvarhelyi2021,Deak2023}, the G center consists of a pair of substituted carbon atoms bonded to an interstitial silicon atom ($\mathrm{C_s-Si_i-C_s}$), and the dipole orientation of G center is in the plane along the [110] direction, as shown in the inset of Fig. 1(a). Fig. 1(b) presents the energy level diagram of the G center, including a ground singlet state, an excited singlet state, and a metastable triplet state \cite{Udvarhelyi2021}. The G center exhibits a ZPL transition at 1278 nm (968 meV), which falls within the telecommunication O-band and can be excited via above-band excitation. As previously mentioned, the CBG cavity offers the advantages of broadband Purcell enhancement, efficient light extraction and collection and considerable tolerance to dipole orientation. In this experiment, the CBG cavity exhibits a quality factor (Q) of approximately 200. Due to the high refractive index contrast between silicon and the surrounding materials, five rings are enough to achieve this Q value. A scanning electron microscope image of a typical CBG cavity is presented in Fig. 1(c). 

\begin{figure*}[htbp]
  \centering
  \includegraphics[width=0.95\textwidth]{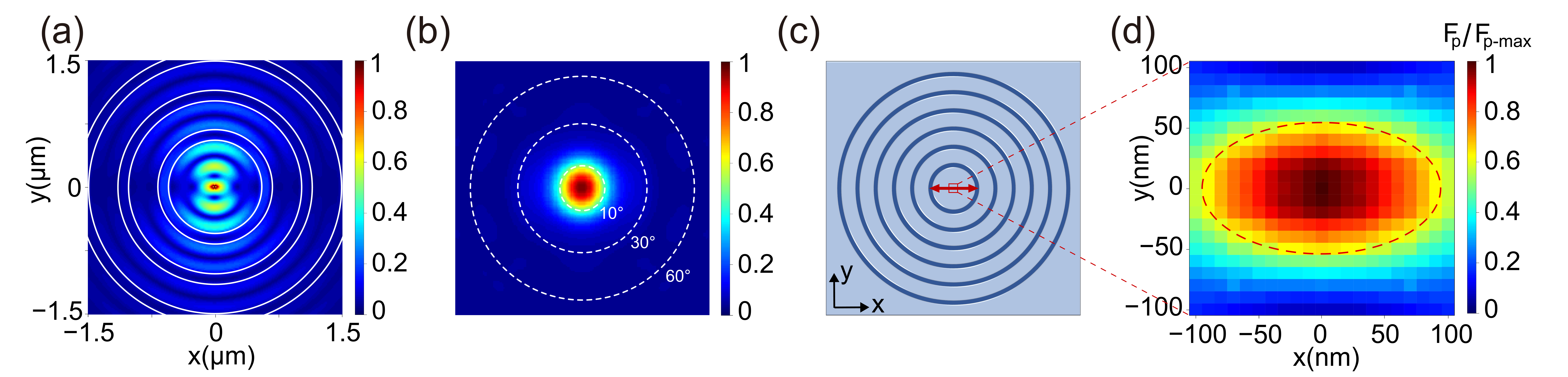}
  \caption{\textbf{Simulation for the single G center integrated in the center of the CBG cavity.} The single G center is modeled as a TE dipole source located at the center of the central disc of the CBG cavity. (a) Top view of the simulated electric field intensity distribution. (b) Simulated far field intensity of the CBG cavity. (c) Schematic of the simulation setup. The TE dipole is in the center of the cavity, along the $x$-direction. (d) Normalized Purcell factors of simulated dipoles located in a 200 nm $\times$ 100 nm grid around the central of the device.}\label{fig2}
\end{figure*}

To elucidate the enhanced luminescence principle of the CBG cavity, we conduct simulations using the finite difference time domain (FDTD) method. The single G center is modeled as a TE dipole source located at the center of the central disc of the CBG cavity. Fig. 2(a) shows a top view of the simulated electric field strength distribution, which indicates that most of the electric field is confined within the central disc. Fig. 2(b) presents the far field radiation map, demonstrating the collimated emission of the dipole source into free space perpendicular to the sample plane, facilitated by the constraints of the CBG cavity. To further investigate the dependence of mode coupling to the CBG cavity on the dipole position, we conduct simulations for a dipole located at different positions relative to the center of the disc. Fig. 2(c) is the schematic of the simulation setup. The resulting normalized Purcell factor showed in Fig. 2(d) indicates that significant Purcell enhancement occurs only within a finite elliptic region of approximately 200 nm $\times$ 100 nm near the center of the disc. It should be emphasized that the observed asymmetry of the dependence on $x$- and $y$- direction originates from the polarized emission of the embedded dipole source \cite{Rickert2019}. The primary mechanism of this enhancement is that the diffraction from the gratings surrounding the disc provide the momentum transfer necessary to couple the in-plane circular cavity modes to the vertical extraction mode \cite{Taillaert2002}. Far field measurement also reveals that the emission is confined within a narrow angle ($\sim10^{\circ}$, see Fig. 2(b)), exhibiting good directivity and significantly improving the collection efficiency of the objective, especially that with low numerical aperture.

Finally, to identify the optimal cavity for coupling the wavelength of the G center ZPL emission with the cavity mode, FDTD simulation and experimental parameter traversal are carried out to obtain the optimal results: diameter of the central disc $d$ = 1100 nm, period width of the surrounding grating ring $l$ = 470 nm, and air gap width $g$ = 125 nm (see Supplementary Note 1 for details).

\begin{figure*}[htbp]
  \centering
  \includegraphics[width=0.95\textwidth]{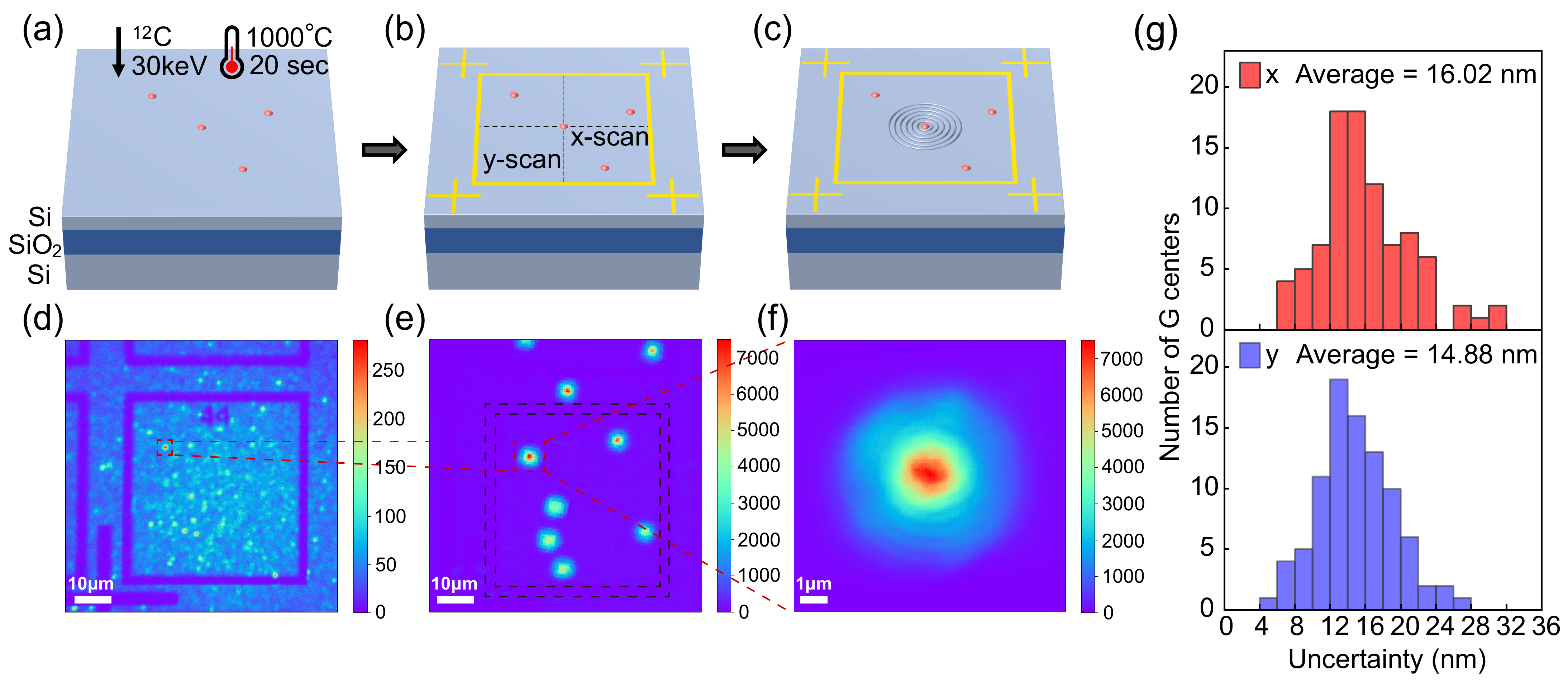}
  \caption{\textbf{Sample fabrication process and confocal PL map of the silicon surface.} (a) Ion implantation and rapid thermal annealing to generate single G centers. (b) Depositing gold markers on the silicon surface to locate the single G center. (c) Manufacturing the CBG cavity at the location of the G center by electron beam exposing through gold alignment markers. (d) PL scan of the sample before cavity fabrication. (e) Confocal PL map after cavity fabrication. (f) High-resolution PL map of the cavity-enhanced G center exhibiting the highest photon count rate in (e). (g) Histograms of the $x$- and $y$-uncertainties of G centers positions, measured by two-dimensional Lorentzian fits to the PL maps from 90 single G centers. The uncertainties represent one standard deviation values determined by a nonlinear least squares fit of the data.}\label{fig3}
\end{figure*}

\textbf{Fluorescence-localization technique (FLT). }In order to deterministically embed a single G center into the center of a CBG cavity, we implement FLT, specifically first locate the single G center based on a fluorescence scan and subsequently employ photolithography and etching to fabricate the cavity. Fig. 3(a)-(c) illustrates the sample fabrication process: (a) The commercial SOI wafer undergoes an ion implantation process, followed by rapid thermal annealing to generate single G centers in the silicon layer (see Sample fabrication in Methods). (b) The pattern of the markers for positioning (yellow crosses and square frames) is formed on the wafer coated with positive Polymethyl Methacrylate (PMMA) resists by photolithography. A 100 nm gold film is coated on the wafer by an e-beam evaporator, and the extra gold is stripped by heating in N-Methylpyrrolidone (NMP) to form the gold markers. The sample is then placed in a cryogenic chamber (7 K), where its surface is scanned using a home-built confocal microscope system (See Supplementary Note 2 for experimental setup details). The typical confocal photoluminescence (PL) map is showed in Fig. 3(d), where the luminous spots correspond to the single G centers while the non-luminous region (purple borders) are the gold markers. Then we use a two-dimensional Lorentzian function to fit the PL maps of the G centers in order to accurately extract the spatial positions of individual G centers. Fig. 3(g) is the histograms of the uncertainties of G centers positions, showing the average uncertainties of $\sim$15 nm (see Supplementary Note 3 for detailed fitting process). (c) The CBG cavities are then deterministically fabricated around the target single G centers by re-exposing the sample with electron beam through the gold alignment markers and subsequently reactive ion etching (see Supplementary Note 4 for detailed sample fabrication process). 

Fig. 3(e) shows the PL map after fabricating the CBG cavities at the previously characterized G center location (corresponding to Fig. 3(d)). To facilitate gold markers identification while maintaining image clarity, gold markers are represented by dotted lines due to their relatively weak emission intensity. A significant PL enhancement can be observed when compared to Fig. 3(d). The high-resolution PL map of the optimally performing CBG cavity exhibiting the highest photon count rate is presented in Fig. 3(f), illustrating a perfect alignment between the single G center and the CBG cavity, which correlates well with the simulated far-field intensity diagram shown in Fig. 2(b). This observation strongly suggests that the G center is situated precisely at the very center of the CBG cavity. The device depicted in Fig. 3(f) is selected for detailed characterization, as described in the following sections.

\begin{figure*}[htbp]
  \centering
  \includegraphics[width=0.8\textwidth]{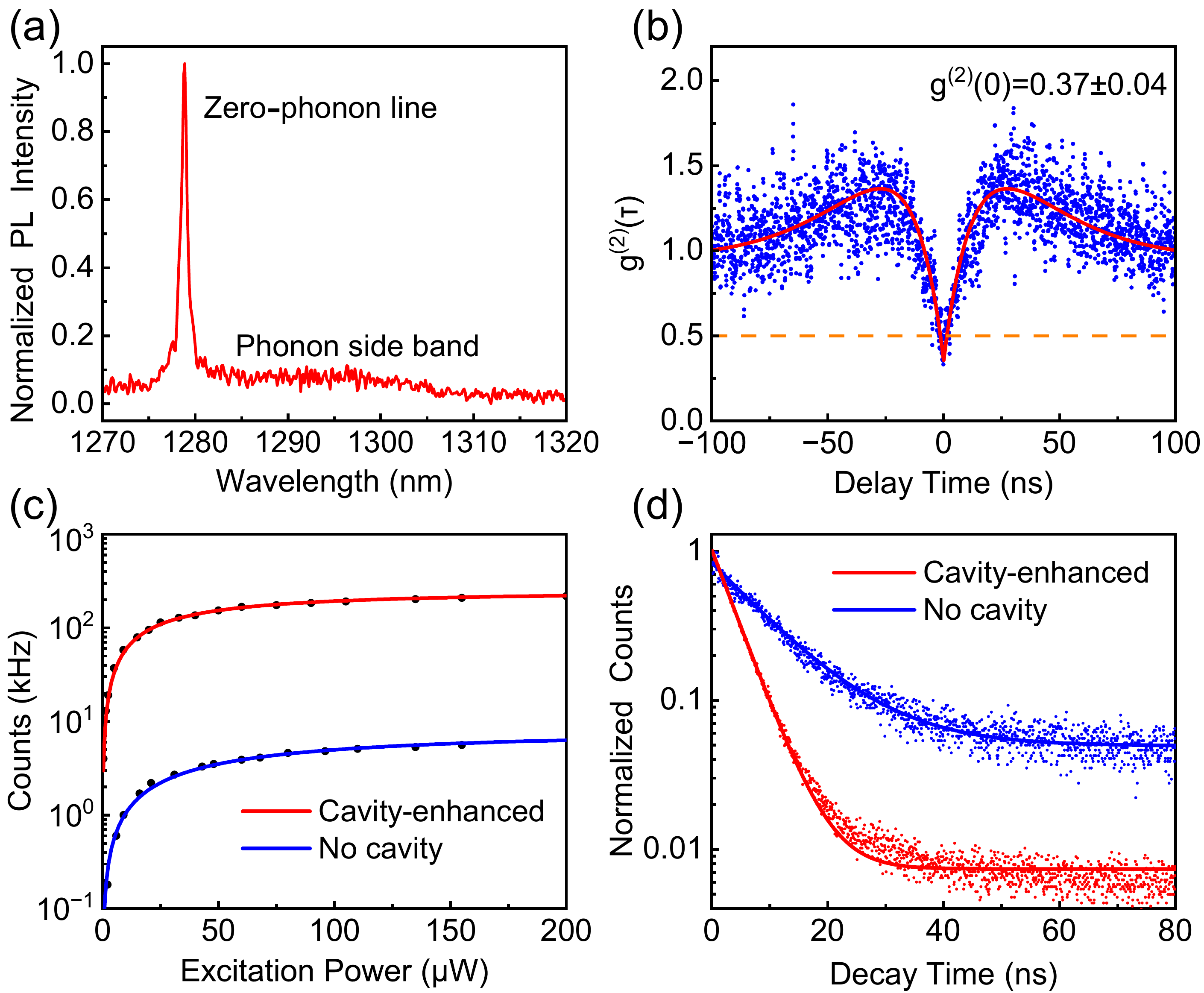}
  \caption{\textbf{Characterization of cavity-enhanced single G center.} (a) PL spectrum of the emitter enhanced by CBG cavity under 532-nm laser excitation at 7 K. (b) Second-order autocorrelation measurement of the G center from the cavity under continuous excitation shows an antibunching at zero delay $g^{(2)}$(0) = 0.37 $\pm$ 0.04. (c) Saturation curve of the G center integrated in a CBG cavity (red curve) and in unpatterned silicon (blue curve) under 532-nm laser excitation. (d) The excited-state lifetime of the G center under pulsed laser excitation. The red and blue curve represent the cavity-enhanced single G center and that without cavity, respectively.}\label{fig4}
\end{figure*}

\textbf{Characterization of cavity-enhanced single G center. }Under 180-$\upmu$W, 532-nm laser excitation, the PL spectrum of the emitter at a low temperature of 7 K is displayed in Fig. 4(a). The spectrum reveals a sharp ZPL located at 1278.8 nm (970.8 meV) along with a broad phonon side band extending to longer wavelengths. To confirm that the bright emission observed within the CBG cavity corresponds to a single G center, we perform second-order autocorrelation measurement for the emitted photons under continuous laser excitation as illustrated in Fig. 4(b) and fit the data using a three-level model:
\begin{equation}
g^{(2)}(\tau)=1-(1-a)\left[(1+b)\text{e}^{-\frac{|t|}{t_1}}-b\text{e}^{-\frac{|t|}{t_2}}\right],
\end{equation}
where $a$ is the normalized zero-delay anti-bunching dip $g^{(2)}$(0), $b$ is a fitting parameter, $t_1$ and $t_2$ are the lifetimes of excited and metastable states, respectively. The resulting second-order correlation of $g^{(2)}$(0) = 0.37 $\pm$ 0.04 reveals an anti-bunching effect, validating that this emitter corresponds to a single G center. It should be noted that this value is predominantly limited by background emission. To investigate further into saturation response characteristics of the single G center, we measure the PL counts under varying laser excitation powers for the single G center coupled to CBG cavity (red line) and single G center without cavity (blue line), as depicted in Fig. 4(c). The collected data are fitted using 
\begin{equation}
I(P)=I_{sat}\cdot\frac{P}{P+P_{sat}},
\end{equation}
which reveals that after coupling into the cavity, the saturation power decreases from $P_{sat1}$ = 72.1 $\upmu$W to $P_{sat2}$ = 34.4 $\upmu$W while the saturation count increases from $I_{sat1}$ = 8.6 kHz to $I_{sat2}$ = 259.3 kHz, showing an improvement factor of approximately 30 times.

We also measure the time-resolved lifetime of the single G center under pulsed laser excitation. The spontaneous emission decay of the cavity-enhanced single G center (red line) and single G center without cavity (blue line) is illustrated in Fig. 4(d). The data are fitted using an exponential decay function. We determine an excited state lifetime of \( t_c = 4.13 \pm 0.01 \, \text{ns} \) for the cavity-coupled single G center, while the lifetime of single G center without cavity is found to be \( 10.27 \pm 0.19 \, \text{ns} \). From these values, we calculate a lifetime enhancement factor \( R = t_c/t_0 = 2.5\).

\section*{Discussions and Conclusion} 
The simulation results presented in Fig. 2(d) reveal a strong spatial dependence of the Purcell enhancement effect on the position of the dipole relative to the cavity center. This spatial sensitivity underscores the critical importance of precise emitter positioning for achieving optimal light-matter interaction. Although focused ion beam (FIB) technology has been previously employed for the controlled fabrication of single G centers \cite{Hollenbach2022_1}, the conventional method involving ion implantation followed by rapid thermal annealing typically yields randomly distributed centers (The FIB method also has its own shortcomings). This spatial randomness results in a low probability of the emission center aligning precisely with the optimal position of the cavity, thereby limiting the achievable enhancement effect.

To overcome this limitation in emitter-cavity alignment, we implement FLT to precisely identify and position individual G centers, ensuring optimal spatial overlap between the emitters and the CBG cavity modes. Unlike traditional methods that involve producing resonators prior to generating SPEs, our approach entails first generating SPEs and precisely mapping color center positions, followed by deterministic cavity patterning. Our technique employs a multi-step process beginning with high-resolution confocal microscopy system to identify and record the coordinates of individual G centers relative to pre-patterned gold alignment markers with $\sim$15 nm spatial precision. These spatial coordinates are then transferred to the electron-beam lithography system, enabling accurate overlay exposure for CBG cavity fabrication.

We selected 40 G centers located by our FLT and fabricated corresponding CBG cavities, 25 of which exhibit significant enhancement in PL intensity, proving a high success probability of 60\%. As for stochastic fabrication approaches, the success probability is only 0.25\%, considering that the density of G centers is 0.04 $\mu$m$^{-2}$ (88 G centers in 46 $\times$ 48-$\mu$m$^{2}$ area) and the effective enhancement area of CBG cavity is 0.063 $\mu$m$^{2}$ (see Fig. 2(d)). This suggests that our FLT represents a 240-times improvement over random match.

By taking advantage of our FLT, we have successfully demonstrated an efficient enhancement of a single G center in silicon by coupling to a CBG cavity. We achieved a remarkable 30-fold enhancement in PL intensity accompanied by a 2.5-fold acceleration of the single-photon emission rate. Next, we estimate the value of the Purcell factor. The Purcell factor is typically defined as the ratio of the radiation lifetime of a cavity-enhanced emitter to that of an emitter without cavities. If the emitter exhibits only radiative transitions, the Purcell factor can be directly expressed as an enhancement in the excited state lifetime. However, it is evident that G center emission involves not only radiative decay but also non-radiative decay and decay into phonon side band, all of which must be considered. Therefore, we refer to Ref. \cite{Islam2024} to give the lower limit of the Purcell factor:
\begin{equation}
F_\mathrm{p} > \frac{R - 1}{D} + 1.
\end{equation}
According to the lifetime enhancement $R$ = 2.5 and the Debye-Waller factor $D$ = 0.15 of G center \cite{Redjem2023}, we get the Purcell factor $F_\mathrm{p} > 11$.

To summarize, we have developed and implemented an advanced FLT that enables accurate positioning and alignment of the quantum emitters within the photonic structure with a high probability of 60\%. The precise spatial alignment between the quantum emitter and cavity, enabled by our deterministic fabrication protocol, represents a significant advancement in silicon-based quantum photonic integration. This achievement not only validates the effectiveness of our localization technique but also establishes a robust platform for scalable quantum photonic circuits in silicon, which is compatible with the mature semiconductor nanofabrication technology today.

\section*{Methods} 
\textbf{Sample fabrication. }Our sample is commercial silicon-on-insulator (SOI) with a 220 nm thick, (100)-oriented, P-type device layer and 3-$\upmu$m thick buried oxide layer. Cleaved chips from this wafer are implanted with $^{12}$C with an energy of 30 keV and a dose of $5\times10^{13}$ cm$^{-2}$. Subsequently, the samples are rapidly thermally annealed at 1000$^{\circ}$C for 20 s in a nitrogen atmosphere to generate single G centers in the silicon layer.

\textbf{Time-resolved lifetime measurements. }Time-resolved lifetime measurements are performed under a 532-nm, 5-MHz pulsed laser excitation coupled with a time-correlated single-photon counting (TCSPC) system. The excitation laser is generated by a white light picosecond pulsed laser (SuperK Ehtreme, NKT Photonics). PL signals are collected and detected using a home-built confocal microscope system. The time-resolved data acquisition is performed using a PicoQuant (HydraHarp), enabling precise measurement of excited-state lifetimes.

\section*{Acknowledgments}
This work is supported by the Innovation Program for Quantum Science and Technology (No. 2021ZD0301200), the National Natural Science Foundation of China (Nos. 124B2082, 12304546, 12174370, 12174376, and 11821404), Anhui Provincial Natural Science Foundation (No. 2308085QA28), China Postdoctoral Science Foundation (No. 2023M733412), the Youth Innovation Promotion Association of Chinese Academy of Sciences (No. 2017492). This work was partially carried out at the USTC Center for Micro and Nanoscale Research and Fabrication.


\begin{thebibliography}{xx}

\bibitem{Bennett2000}
Bennett, C. H. \& DiVincenzo, D. P. Quantum information and computation. \href{https://www.nature.com/articles/35005001}{\textcolor{blue}{\emph{Nature} \textbf{404}, 247-255 (2000)}}.


\bibitem{Gisin2007}
Gisin, N. \& Thew, R. Quantum communication. \href{https://www.nature.com/articles/nphoton.2007.22}{\textcolor{blue}{\emph{Nat. Photon.} \textbf{1}, 165-171 (2007)}}.

\bibitem{Nielsen2010}
Nielsen, M. A. \& Chuang, I. L. Quantum computation and quantum information: 10th anniversary edition. \href{https://www.cambridge.org/highereducation/books/quantum-computation-and-quantum-information/01E10196D0A682A6AEFFEA52D53BE9AE#overview}{\textcolor{blue}{\emph{Cambridge University Press} (2010)}}.

\bibitem{Degen2017}
Degen, C. L., Reinhard, F. \& Cappellaro, P. Quantum sensing. \href{https://journals.aps.org/rmp/abstract/10.1103/RevModPhys.89.035002}{\textcolor{blue}{\emph{Rev. Mod. Phys.} \textbf{89}, 035002 (2017)}}.

\bibitem{Doherty2013}
Doherty, M. W. \emph{et al}. The nitrogen-vacancy colour centre in diamond. \href{https://www.sciencedirect.com/science/article/pii/S0370157313000562}{\textcolor{blue}{\emph{Phys. Rep.} \textbf{528}, 1-45 (2013)}}.

\bibitem{Hepp2014}
Sipahigil, A. \emph{et al}. Indistinguishable photons from separated silicon-vacancy centers in diamond. \href{https://journals.aps.org/prl/abstract/10.1103/PhysRevLett.113.113602}{\textcolor{blue}{\emph{Phys. Rev. Lett.} \textbf{113}, 113602 (2014)}}.

\bibitem{Lohrmann2017}
Lohrmann, A., Johnson, B. C., McCallum, J. C. \& Castelletto, S. A review on single photon sources in silicon carbide. \href{https://iopscience.iop.org/article/10.1088/1361-6633/aa5171}{\textcolor{blue}{\emph{Rep. Prog. Phys.} \textbf{80}, 034502 (2017)}}.

\bibitem{Wang2019}
Wang, J.-F. \emph{et al}. On-demand generation of single silicon vacancy defects in silicon carbide. \href{https://pubs.acs.org/doi/10.1021/acsphotonics.9b00451}{\textcolor{blue}{\emph{ACS Photon.} \textbf{6}, 1736-1743 (2019)}}.

\bibitem{Guo2023}
Guo, N. J. \emph{et al}. Coherent control of an ultrabright single spin in hexagonal boron nitride at room temperature. \href{https://www.nature.com/articles/s41467-023-38672-6}{\textcolor{blue}{\emph{Nat. Commun.} \textbf{4}, 552-567 (2019)}}.

\bibitem{Silverstone2016}
Silverstone, J. W., Bonneau, D., O' Brien, J. L. \& Thompson, M. G. Silicon quantum photonics. \href{https://ieeexplore.ieee.org/document/7479523}{\textcolor{blue}{\emph{IEEE J. Sel. Top. Quantum Electron.} \textbf{22}, 390-402 (2016)}}.

\bibitem{Samkharadze2018}
Samkharadze, N. \emph{et al}. Strong spin-photon coupling in silicon. \href{https://www.science.org/doi/10.1126/science.aar4054}{\textcolor{blue}{\emph{Science} \textbf{359}, 1123-1127(2018)}}.

\bibitem{Yin2013}
Yin, C. \emph{et al}. Optical addressing of an individual erbium ion in silicon. \href{https://www.nature.com/articles/nature12081}{\textcolor{blue}{\emph{Nature} \textbf{497}, 91-94 (2013)}}.

\bibitem{Raha2020}
Raha, M. \emph{et al}. Optical quantum nondemolition measurement of a single rare earth ion qubit. \href{https://www.nature.com/articles/s41467-020-15138-7}{\textcolor{blue}{\emph{Nat. Commun.} \textbf{11}, 1605 (2020)}}.

\bibitem{Chen2020}
Chen, S., Raha, M., Phenicie, C. M., Ourari, S. \& Thompson, J. D. Parallel single-shot measurement and coherent control of solid-state spins below the diffraction limit. \href{https://www.science.org/doi/10.1126/science.abc7821}{\textcolor{blue}{\emph{Science} \textbf{370}, 592-595 (2020)}}.

\bibitem{Ourari2023}
Ourari, S. \emph{et al}. Indistinguishable telecom band photons from a single Er ion in the solid state. \href{https://www.nature.com/articles/s41586-023-06281-4}{\textcolor{blue}{\emph{Nature} \textbf{620}, 977-981 (2023)}}.

\bibitem{Khoury2022}
Khoury, M. \& Abbarchi, M. A bright future for silicon in quantum technologies. \href{https://pubs.aip.org/aip/jap/article/131/20/200901/2836887}{\textcolor{blue}{\emph{J. Appl. Phys.} \textbf{131}, 200901 (2022)}}.

\bibitem{Durand2021}
Durand, A. et al. Broad diversity of near-infrared single-photon emitters in silicon. \href{https://journals.aps.org/prl/abstract/10.1103/PhysRevLett.126.083602}{\textcolor{blue}{\emph{Phys. Rev. Lett.} \textbf{126}, 083602 (2021)}}.

\bibitem{Redjem2020}
Redjem, W. et al. Single artificial atoms in silicon emitting at telecom wavelengths. \href{https://www.nature.com/articles/s41928-020-00499-0}{\textcolor{blue}{\emph{Nat. Electron.} \textbf{3}, 738-743 (2020)}}.

\bibitem{Hollenbach2020}
Hollenbach, M., Berenc\'en, Y., Kentsch, U., Helm, M. \& Astakhov, G. V. Engineering telecom single-photon emitters in silicon for scalable quantum photonics. \href{https://opg.optica.org/oe/fulltext.cfm?uri=oe-28-18-26111&id=437438}{\textcolor{blue}{\emph{Opt. Express} \textbf{28}, 26111 (2020)}}.

\bibitem{Udvarhelyi2021}
Udvarhelyi, P., Somogyi, B., Thiering, G. \& Gali, A. Identification of a Telecom Wavelength Single Photon Emitter in Silicon. \href{https://journals.aps.org/prl/abstract/10.1103/PhysRevLett.127.196402}{\textcolor{blue}{\emph{Phys. Rev. Lett.} \textbf{127}, 196402 (2021)}}.

\bibitem{Baron2022_1}
Baron, Y. \emph{et al}. Single G centers in silicon fabricated by co-implantation with carbon and proton. \href{https://pubs.aip.org/aip/apl/article/121/8/084003/2834250}{\textcolor{blue}{\emph{Appl. Phys. Lett.} \textbf{121}, 084003 (2022)}}.

\bibitem{Hollenbach2022_1}
Hollenbach, M. \emph{et al}. Wafer-scale nanofabrication of telecom single-photon emitters in silicon. \href{https://www.nature.com/articles/s41467-022-35051-5}{\textcolor{blue}{\emph{Nat. Commun.} \textbf{13}, 7683 (2022)}}.

\bibitem{Prabhu2023}
Prabhu, M. \emph{et al}. Individually addressable and spectrally programmable artificial atoms in silicon photonics. \href{https://www.nature.com/articles/s41467-023-37655-x}{\textcolor{blue}{\emph{Nat. Commun.} \textbf{14}, 2380 (2023)}}.

\bibitem{Higginbottom2022}
Higginbottom, D. B. \emph{et al}. Optical observation of single spins in silicon. \href{https://www.nature.com/articles/s41586-022-04821-y}{\textcolor{blue}{\emph{Nature} \textbf{607}, 266-270 (2022)}}.

\bibitem{Baron2022_2}
Baron, Y. \emph{et al}. Detection of single W-centers in silicon. \href{https://pubs.acs.org/doi/10.1021/acsphotonics.2c00336}{\textcolor{blue}{\emph{ACS Photon.} \textbf{9}, 2337-2345 (2022)}}.

\bibitem{Johnston2024}
Johnston, A., Felix-Rendon, U., Wong, Y.-E. \& Chen, S. Cavity-coupled telecom atomic source in silicon. \href{https://www.nature.com/articles/s41467-024-46643-8}{\textcolor{blue}{\emph{Nat. Commun.} \textbf{15}, 2350 (2024)}}.

\bibitem{Islam2024}
Islam, F. \emph{et al}. Cavity-enhanced emission from a silicon T center. \href{https://pubs.acs.org/doi/10.1021/acs.nanolett.3c04056}{\textcolor{blue}{\emph{Nano Lett.} \textbf{24}, 319-325 (2024)}}.

\bibitem{Redjem2023}
Redjem, W. \emph{et al}. All-silicon quantum light source by embedding an atomic emissive center in a nanophotonic cavity. \href{https://www.nature.com/articles/s41467-023-38559-6}{\textcolor{blue}{\emph{Nat. Commun.} \textbf{14}, 3321 (2023)}}.

\bibitem{Saggio2024}
Saggio, V. \emph{et al}. Cavity-enhanced single artificial atoms in silicon. \href{https://www.nature.com/articles/s41467-024-49302-0}{\textcolor{blue}{\emph{Nat. Commun.} \textbf{15}, 5296 (2024)}}.

\bibitem{Komza2024}
Komza, L. \emph{et al}. Indistinguishable photons from an artificial atom in silicon photonics. \href{https://www.nature.com/articles/s41467-024-51265-1}{\textcolor{blue}{\emph{Nat. Commun.} \textbf{15}, 6920 (2024)}}.

\bibitem{Tait2020}
Tait, A. N. \emph{et al}. Microring resonator-coupled photoluminescence from silicon W centers. \href{https://iopscience.iop.org/article/10.1088/2515-7647/ab95f2/meta}{\textcolor{blue}{\emph{J. Phys. Photonics} \textbf{2}, 045001 (2020)}}.

\bibitem{Lefaucher2023}
Lefaucher, B. \emph{et al}. Cavity-enhanced zero-phonon emission from an ensemble of G centers in a silicon-on-insulator microring. \href{https://pubs.aip.org/aip/apl/article/122/6/061109/2867575}{\textcolor{blue}{\emph{Appl. Phys Lett.} \textbf{122}, 061109 (2023)}}.

\bibitem{Zhu2020}
Zhu, L., Yuan, S., Zeng, C. \& Xia, J. Manipulating Photoluminescence of carbon G-center in silicon metasurface with optical bound states in the continuum. \href{https://onlinelibrary.wiley.com/doi/10.1002/adom.201901830}{\textcolor{blue}{\emph{Adv. Opt. Mater.} \textbf{8}, 1901830 (2020)}}.

\bibitem{Hollenbach2022_2}
Hollenbach, M. \emph{et al}. Metal-assisted chemically etched silicon nanopillars hosting telecom photon emitters. \href{https://pubs.aip.org/aip/jap/article/132/3/033101/2837082}{\textcolor{blue}{\emph{J. Appl. Phys.} \textbf{132}, 033101 (2022)}}.

\bibitem{Lefaucher2024}
Lefaucher, B. \emph{et al}. Purcell enhancement of silicon W centers in circular Bragg grating cavities. \href{https://pubs.acs.org/doi/10.1021/acsphotonics.3c01561}{\textcolor{blue}{\emph{ACS Photon.} \textbf{11}, 24-32 (2024)}}.

\bibitem{Rickert2019}
Rickert, L., Kupko, T., Rodt, S., Reitzenstein, S. \& Heindel, T. Optimized designs for telecom-wavelength quantum light sources based on hybrid circular Bragg gratings. \href{https://opg.optica.org/oe/fulltext.cfm?uri=oe-27-25-36824&id=423857}{\textcolor{blue}{\emph{Opt. Express} \textbf{27}, 36824-36837 (2019)}}.

\bibitem{Li2015}
Li, L. \emph{et al}. Efficient photon collection from a nitrogen vacancy center in a circular bullseye grating. \href{https://pubs.acs.org/doi/10.1021/nl503451j}{\textcolor{blue}{\emph{Nano Lett.} \textbf{15}, 1493-1497 (2015)}}.

\bibitem{Froch2021}
Fr$\rm \ddot{o}$ch, J. E. \emph{et al}. Coupling spin defects in hexagonal boron nitride to monolithic bullseye cavities. \href{https://pubs.acs.org/doi/10.1021/acs.nanolett.1c01843}{\textcolor{blue}{\emph{Nano Lett.} \textbf{21}, 6549-6555 (2021)}}.

\bibitem{Liu2019}
Liu, J. \emph{et al}. A solid-state source of strongly entangled photon pairs with high brightness and indistinguishability. \href{https://www.nature.com/articles/s41565-019-0435-9}{\textcolor{blue}{\emph{Nat. Nanotech.} \textbf{14}, 586-593 (2019)}}.

\bibitem{Kolatschek2021}
Kolatschek, S. \emph{et al}. Bright Purcell enhanced single-Photon source in the telecom O-band based on a quantum dot in a circular Bragg grating. \href{https://pubs.acs.org/doi/10.1021/acs.nanolett.1c02647}{\textcolor{blue}{\emph{Nano Lett.} \textbf{21}, 7740-7745 (2021)}}.

\bibitem{Deak2023}
De$\acute{a}$k, P., Udvarhelyi, P., Thiering, G. \& Gali, A. The kinetics of carbon pair formation in silicon prohibits reaching thermal equilibrium. \href{https://www.nature.com/articles/s41467-023-36090-2}{\textcolor{blue}{\emph{Nat. Commun.} \textbf{14}, 361 (2023)}}.

\bibitem{Taillaert2002}
Taillaert, D. \emph{et al}. An out-of-plane grating coupler for efficient butt-coupling between compact planar waveguides and single-mode fibers. \href{https://ieeexplore.ieee.org/document/1017613}{\textcolor{blue}{\emph{IEEE J. Quantum Electron.} \textbf{38}, 949-955 (2002)}}.


\end{thebibliography}
\end{document}